\documentclass[11pt]{article}
\usepackage{xypic}
\usepackage[color,matrix,arrow]{xy}
\usepackage{setspace}

\usepackage{amsmath,amssymb,graphicx,pict2e}
\usepackage{hyperref}
\usepackage{multicol,color,longtable}
\definecolor{darkred}{rgb}{0.65,0.15,0}
\hypersetup{pdfborder={0 0 0},colorlinks=true,urlcolor=blue,citecolor=blue,linkcolor=darkred,linktocpage=true}

\usepackage{bbold}

\usepackage{cite}
\usepackage{array,multirow,stmaryrd}

\usepackage{young}
\usepackage[vcentermath]{youngtab}

\newcommand{\eprint}[1]{{\href{http://arxiv.org/abs/#1}{[\texttt{#1}]}}}
\newcommand{\eprintN}[1]{{\href{http://arxiv.org/abs/#1}{[\texttt{#1 [hep-th]}]}}}
\newcommand{\eprintMPH}[1]{{\href{http://arxiv.org/abs/#1}{[\texttt{#1 [math-ph]}]}}}
\newcommand{\eprintgrqc}[1]{{\href{http://arxiv.org/abs/#1}{[\texttt{#1 [gr-qc]}]}}}
\newcommand{\eprintgrqcold}[1]{{\href{http://arxiv.org/abs/#1}{[\texttt{#1}]}}}
\newcommand{\doi}[2]{{\hypersetup{urlcolor=darkred}\href{http://dx.doi.org/#2}{#1}\hypersetup{urlcolor=blue}}}

\newcommand{\nn}{\nonumber}

\newcommand{\reals}{\mathbb{R}}

\newcommand{\lra}{\leftrightarrow}

\newcommand{\mf}[1]{{\mathfrak{#1}}}

\newcommand{\lb}{\left[}
\newcommand{\rb}{\right]}
\newcommand{\lp}{\left(}
\newcommand{\rp}{\right)}

\newcommand{\ALT}{\textrm{\large{$\wedge$}}}

\newcommand{\Minfty}{\textrm{Maxwell}_{\infty}}

\makeatletter

\@addtoreset{equation}{section} \makeatother

\begin{document}

{\flushright {ICCUB-19-010}\\[5mm]}

\begin{center}
{\LARGE \bf Galilean Free Lie Algebras}\\[10mm]

\vspace{8mm}
\normalsize
{\large  Joaquim Gomis${}^{1}$, Axel Kleinschmidt${}^{2,3}$ and Jakob Palmkvist${}^4$}

\vspace{10mm}
${}^1${\it Departament de F\'isica Qu\`antica i Astrof\'isica\\ and
Institut de Ci\`encies del Cosmos (ICCUB), Universitat de Barcelona\\ Mart\'i i Franqu\`es , ES-08028 Barcelona, Spain}
\vskip 1 em
${}^2${\it Max-Planck-Institut f\"{u}r Gravitationsphysik (Albert-Einstein-Institut)\\
Am M\"{u}hlenberg 1, DE-14476 Potsdam, Germany}
\vskip 1 em
${}^3${\it International Solvay Institutes\\
ULB-Campus Plaine CP231, BE-1050 Brussels, Belgium}
\vskip 1 em
${}^4${\it Division for Theoretical Physics, Department of Physics, Chalmers University of Technology\\ SE-412 96 Gothenburg, Sweden}

\vspace{20mm}

\hrule

\vspace{10mm}

\begin{tabular}{p{12cm}}
{\small
We construct free Lie algebras which, together with the algebra of spatial rotations, form infinite-dimensional extensions of
finite-dimensional Galilei Maxwell algebras appearing as global spacetime symmetries of
extended non-relativistic objects and non-relativistic gravity theories. We show how various extensions of the ordinary Galilei algebra can be obtained by truncations and contractions, in some cases via an affine Kac--Moody algebra. The infinite-dimensional Lie algebras could be useful in the construction of
generalized Newton--Cartan theories gravity theories and the objects that couple to them.
}
\end{tabular}
\vspace{7mm}
\hrule
\end{center}

\newpage

\setcounter{tocdepth}{2}
\tableofcontents

\vspace{5mm}
\hrule
\vspace{5mm}

\section{Introduction and summary}

Relativistic particles coupled to a constant electro-magnetic field enjoy symmetries that extend the usual Poincar\'e algebra symmetries~\cite{Bacry:1970ye,Schrader:1972zd}. Assuming that the 
electro-magnetic field transforms, one obtains a non-central extension of the Poincar\'e algebra by an anti-symmetric tensor generator $Z_{ab}$ that transforms covariantly under the Lorentz group and the resulting algebra has been called the Maxwell algebra in~\cite{Schrader:1972zd}. This algebra has also been recovered from studying the Chevalley--Eilenberg cohomology of the Poincar\'e algebra~\cite{Bonanos:2008kr,Bonanos:2008ez}, where also further non-central extensions have been identified. The algebraic structure can be embedded in a free Lie algebra construction as shown in~\cite{Gomis:2017cmt} such that different quotients of the free Lie algebra yield the known relativistic Maxwell algebra or related extensions~\cite{Beckers:1983gp,Soroka:2004fj,Gomis:2009dm,Salgado:2014qqa}. A similar analysis was undertaken for the supersymmetric 
theory in~\cite{Bonanos:2009wy,Concha:2018jxx,Penafiel:2017wfr,Ravera:2018vra,Andrianopoli:2016osu,Gomis:2018xmo}.

For massive non-relativistic particles the algebra obtained by contraction of the Poincar\'e algebra is the Galilei algebra.  A  non-relativistic Maxwell-type extension in a constant electro-magnetic background is known to arise~\cite{Bacry:1970du,Beckers:1983gp} and this has also been understood from Lie algebra cohomology~\cite{Bonanos:2008kr}. However, in the non-relativistic case there are more options depending on how the contracting limit  \cite{lebellac,Barducci:2019fjc,Duval:1990hj} is taken from the relativistic Maxwell algebra. 

In the present paper we shall take the different non-relativistic limits of the Maxwell algebra as a starting point and consider their embedding in free Lie algebra constructions. As was investigated in~\cite{Barducci:2019fjc} and as we shall explain in more detail in section~\ref{sec:GFLA},  
there are three distinct non-relativistic limits of the Maxwell algebra in the point particle case depending on the relative scaling between the electric and magnetic field as the speed of light is sent to infinity. 
These three cases can be embedded in free Lie algebra extensions and two of them interestingly require different free Lie algebra constructions.
The free Lie algebra permit a plethora of quotients that connect to the Bargmann algebra and other non-relativistic symmetry algebras that have appeared recently in the literature~\cite{Hansen:2018ofj,Ozdemir:2019orp,Hansen:2019vqf,Bergshoeff:2018vfn}.

Investigating the possible non-relativistic symmetry algebras is relevant for obtaining physical models with non-relativistic invariances. Beyond point particle dynamics, one can also consider realisations that lead to non-relativistic 
gravity~\cite{Cartan1,Cartan2,Trautman63,Havas:1964zza,DePietri:1994je, 
Andringa:2010it, Aviles:2018jzw,Hansen:2018ofj,Ozdemir:2019orp,Hansen:2019vqf,
Bergshoeff:2019ctr,
deAzcarraga:2019mdn,Concha:2019lhn,Penafiel:2019czp} or even symmetries of extended objects such as strings~\cite{Gomis:2000bd,Danielsson:2000gi,Brugues:2004an} branes~\cite{Brugues:2006yd,
Batlle:2016iel}
and stringy non-relativistic gravities \cite{Andringa:2012uz,Bergshoeff:2018vfn,
Aviles:2019xed}

In this paper we shall not study the dynamical realisations of the symmetry algebras we present. However, we expect that there can be additional symmetries in a given dynamical realisation beyond the one that was used when writing down the 
model~\cite{Niederer:1972zz,Hagen:1972pd}, see also~\cite{Batlle:2016iel}. This happens for example already for the non-relativistic massive particle that enjoys Schr\"odinger 
\cite{Niederer:1972zz,Hagen:1972pd} invariance even though the original symmetry algebra is only Galilei as can be seen by studying the Killing symmetries of the model. The Schr\"odinger algebra is the maximal point symmetry group  it has three extra generators with respect to the Galilei algebra, these are the central charge associated to the mass of the particle, the 
dilatation $D$ that gives a different scaling for the time coordinates and  spatial coordinates and the one-dimensional special conformal transformation
$C$.\footnote{The generators of time translation $H$, dilatation $D$ and one-dimensional
special conformal transformations $C$ form an $SL(2,\reals)$ algebra.}
It is not clear how to fit this extended algebra into a free Lie algebra construction similar to the one considered in the present paper.

One of the features of our analysis is that it also provides a bridge to (affine) Kac--Moody algebras and so-called Lie algebra expansions. 
The latter were studied in~\cite{deAzcarraga:2002xi,Izaurieta:2006zz,deAzcarraga:2007et} in order
 to construct a series of new Lie algebras from an initial given Lie algebra. The method has also been used
as a way of obtaining actions that are invariant under an extended (``expanded'') symmetry algebra when starting from an action with a given smaller symmetry algebra. In~\cite{Bergshoeff:2019ctr} it was applied systematically in order to obtain various non-relativistic gravitational models in $D=3$ and $D=4$ from the Einstein--Hilbert action by expanding the Poincar\'e algebra arranged in a non-relativistic point particle split. As we shall explain in section~\ref{sec:AKM}, the expansion method is the same as constructing an affine Kac--Moody extension of the finite-dimensional Lorentz algebra and considering a truncation and contraction of the Borel subalgebra. The connection can be traced back to the fact that both the affine algebra and the expanded Lie algebra can be thought of as formal power series in an expansion parameter.

The reason for the greater variety of non-relativistic symmetries compared to the relativistic case can be traced back to the fact that the 
 non-relativistic space and time coordinates can have a different scaling. There
 are (at least) three different ways of constructing models with Galilean invariance: (i) from non-relativistic limits of relativistic theories  
\cite{Havas:1964zza,Bacry:1968zf,Dautcourt:1990,Dautcourt:1996pm,Gomis:2000bd,Batlle:2016iel,VandenBleeken:2017rij}, 
(ii) doing a null reduction of relativistic theories in one dimension higher 
\cite{Eisenhart,Gomis:1978mv,Duval:1984cj,Julia:1994bs} and
(iii) an {\it ab initio} non-relativistic construction \cite{Cartan1,Cartan2,Trautman63,bargmann, LevyLeblond:1967zz,Sou,Niederle:2008zza}. As mentioned above, we do not consider any specific non-relativistic model in the present paper but rather present possible algebraic building blocks for Galilean invariance of point particles or extended objects that we hope prove useful for explicit constructions. Such explicit models in turn are important for understanding better non-relativistic gravity, in particular in the context of non-relativistic holography~\cite{Sachdev,Liu} or non-relativistic string theory~\cite{Gomis:2000bd,Danielsson:2000gi,Batlle:2016iel,
Bergshoeff:2018yvt,Gomis:2019zyu,Gallegos:2019icg}.

This paper is structured as follows. In section~\ref{sec:GFLA}, we recall three distinct contractions of the relativistic Maxwell algebra~\cite{lebellac,Barducci:2019fjc}
and study how the resulting three extensions of the Galilei algebra
in the point particle case
 are embedded in distinct free Lie algebras. We discuss how their quotients give rise to known algebras. Section~\ref{sec:AKM} is devoted to studying a particularly interesting case that has recently attracted attention in connection with non-relativistic gravity~\cite{Hansen:2018ofj,Ozdemir:2019orp,Hansen:2019vqf,
Bergshoeff:2019ctr,deAzcarraga:2019mdn,Concha:2019lhn,Penafiel:2019czp
} and we show that this algebra can actually be obtained as a truncation of an affine Kac--Moody algebra. In section~\ref{sec:SGA}, we generalise from the Galilei algebra of point particles to that of extended $p$-branes. Focussing our attention on one particular embedding in a free Lie algebra we recover known central and non-central extensions of the string Galilei algebra of~\cite{Brugues:2004an}. 

\subsection*{Acknowledgements}

We acknowledge discussions with Luis Avil\'es, Andrea Barducci, Eric Bergshoeff, Roberto Casalbuoni, Jaume Gomis, Tomas Ort\'{i}n, Patricio Salgado-Rebolledo and Jakob Salzer.
AK gratefully acknowledges the hospitality of the University of Barcelona and of Chalmers University of Technology, and JG and JP
the hospitality of the Albert Einstein Institute.
JG has been supported in part by MINECO FPA2016-76005-C2-1-P and Consolider CPAN, and by the Spanish government (MINECO/FEDER) under project MDM-2014-0369 of ICCUB (Unidad de Excelencia Mar\'{i}a de Maeztu). JP has been supported by the Swedish Research Council, project no.\ 2015-04268.

\section{Galilean free Lie algebras}
\label{sec:GFLA}

We consider the question of embedding Galilean algebras in free Lie algebras in a fashion similar to the embedding of the (supersymmetric) Maxwell algebra in a free Lie \mbox{(super)}algebra~\cite{Gomis:2017cmt,Gomis:2018xmo}. Before considering the construction of the free Lie algebra, we first study the different (unextended) Galilean algebras that can arise in an electro-magnetic context.

\subsection{Galilei Maxwell algebras}

Going back to Le Bellac and Levy-Leblond one can study several limits of the relativistic Maxwell and Lorentz equations that lead to different forms of non-relativistic systems. In the original paper~\cite{lebellac}, two different forms of Galilean electromagnetisms were constructed:
\begin{enumerate}
\item[1)] The \textit{magnetic} Galilean Maxwell equations, where the non-relativistic limit was taken with the magnetic field much larger than the electric field.
\item[2)] The \textit{electric} Galilean Maxwell equations, where the limit was taken with the electric field much larger than the magnetic one.
\end{enumerate}  
One could also consider a third case:
\begin{enumerate}
\item[3)] The \textit{pulse/shockwave} Galilean Maxwell equations, where the limit is taken with the electric field and magnetic field large and of equal modulus.
\end{enumerate}

This last case has also appeared in a recent study~\cite{Barducci:2019fjc} of contractions of the Maxwell algebra. The Maxwell algebra~\cite{Schrader:1972zd} is a tensorial extension of the Poincar\'e algebra of the form $\lb P_a, P_b \rb=Z_{ab}$ with a new anti-symmetric Lorentz tensor generator $Z_{ab}$ that is associated with a constant electro-magnetic field. This algebra and its realisations have been further studied, see for example \cite{Bonanos:2008ez,Gomis:2017cmt}.  

The magnetic limit of the Maxwell algebra in $D$ space-time dimensions can be obtained by considering the generators
\begin{align}
\label{eq:contr}
\tilde{M}_{ij} &= M_{ij}\,,& \tilde{G}_i &= \frac1{\omega} M_{i0}\,,\nn\\
\tilde{H} &= \omega P_0\,, & \tilde{P}_i &= P_i\,,\\
\tilde{Z}_{ij} &=  Z_{ij} \,,& \tilde{Z}_i &= \omega Z_{0i}\nn
\end{align}
and sending $\omega\to \infty$. \textit{In order to keep the notation light, we shall henceforth drop the tildes on the generators after contraction.} Our indexing notation above is such that the spatial indices $i$ run from $1$ to $D-1$. We refer to $M_{ij}$ as the spatial rotations, $G_i$ as the boost generators, $H$ as the Hamiltonian, $P_i$ as the spatial (transverse) translations, $Z_{ij}$ as the magnetic field and $Z_i$ as the electric field. The algebra obtained from the relativistic Maxwell algebra after the magnetic contraction above is\footnote{We are using the conventions $[M_{ab}, M_{cd}] = \eta_{bc} M_{ad} +\ldots$ and $[M_{ab}, P_c] = \eta_{bc} P_a - \eta_{ac} P_b$ for the relativistic Poincar\'e algebra with mostly plus Minkowski metric as in previous work~\cite{Gomis:2017cmt}.}
\begin{align} 
\label{eq:MGM}
 \lb G_i, P_j\rb &= 0\,,&   \lb M_{ij}, P_k\rb  &=   2\delta_{k[j} P_{i]}  \,,& \lb G_i, Z_j\rb &= -Z_{ij} \,, \nn\\ 
 \lb H,G_i\rb  &=  P_i \,,& \lb M_{ij},G_k \rb &=  2 \delta_{k[j} G_{i]} \,,&  \lb P_i, P_j\rb &= Z_{ij} \,,\\
  \lb H, P_i\rb &= Z_i \,,& \lb M_{ij}, Z_k\rb &=  2\delta_{k[j}Z_{i]} \,,&  \lb  G_k, Z_{ij}  \rb &=0\,,  \nn\\
\lb G_i, G_j\rb &= 0\,, & \lb M_{ij} , Z_{kl} \rb &= -4 \delta_{[i[l} Z_{k]j]}\,.\nn
\end{align}
In the terminology of~\cite{Barducci:2019fjc} this is the $k=1$ contraction of the Maxwell algebra, corresponding to a point particle. In section~\ref{sec:SGA}, we shall consider generalisations to extended objects.

By contrast, the electric Maxwell algebra is (this differs by scaling the magnetic field $Z_{ij}$ by $\omega^2$ instead of $\omega^0$)
\begin{align} 
\label{eq:EGM}
 \lb G_i, P_j\rb &= 0\,,& \lb M_{ij}, P_k\rb  &=   2\delta_{k[j} P_{i]}  \,,&   \lb G_i,  Z_j\rb &= 0 \,, \nn\\ 
 \lb H, G_i\rb  &=  P_i \,,& \lb  M_{ij}, G_k \rb &=  2 \delta_{k[j}  G_{i]} \,,&  \lb P_i, P_j\rb &= 0\,,\\
  \lb H, P_i\rb &=  Z_i \,,& \lb M_{ij}, Z_k\rb &=  2\delta_{k[j} Z_{i]}\,,&  \lb   G_k, Z_{ij}  \rb &=2\delta_{k[i} Z_{j]}\,,  \nn\\
\lb G_i, G_j\rb &= 0\,, & \lb M_{ij} , Z_{kl} \rb &= -4 \delta_{[i[l} Z_{k]j]}\,.\nn
\end{align}
Comparing equations~\eqref{eq:MGM} and~\eqref{eq:EGM} we see that the difference lies in the third columns of equations. For instance in the electric limit~\eqref{eq:EGM} the large electric field $Z_i$ becomes boost-invariant while it is the large magnetic field $Z_{ij}$ that is boost-invariant in~\eqref{eq:MGM}. Moreover, the non-trivial commutator $[P_i, P_j]$ in~\eqref{eq:MGM} is inherited from the Maxwell algebra while it is contracted to zero in the electric case.

We also note that the usual unextended Galilei algebra is obtained from either Maxwell Galilei algebra by setting to zero the electro-magnetic generators $Z_i$ and $Z_{ij}$.

In the pulse case the algebra becomes (the scaling for $Z_{ij}$ is with $\omega$ in this case, so that the electric and magnetic fields
scale in the same way)~\cite{Barducci:2019fjc}
\begin{align} 
\label{eq:PGM}
 \lb G_i,P_j\rb &= 0\,,& \lb M_{ij}, P_k\rb  &=   2\delta_{k[j} P_{i]}  \,,&   \lb G_i, Z_j\rb &= 0 \,, \nn\\ 
 \lb H, G_i\rb  &=  P_i \,,& \lb M_{ij}, G_k \rb &=  2 \delta_{k[j} G_{i]} \,,&  \lb P_i, P_j\rb &= 0\,,\\
  \lb H, P_i\rb &= Z_i \,,& \lb M_{ij}, Z_k\rb &=  2\delta_{k[j} Z_{i]}\,,&  \lb  G_k, Z_{ij}  \rb &=0\,,  \nn\\
\lb G_i,G_j\rb &= 0\,, & \lb M_{ij} , Z_{kl} \rb &= -4 \delta_{[i[l} Z_{k]j]}\,.\nn
\end{align}
Comparing this to~\eqref{eq:MGM} and~\eqref{eq:EGM} we see that both the large electric and magnetic field generators have become boost invariant and that the magnetic field generator is not generated by the commutator of two translations anymore.

We note that the pulse algebra~\eqref{eq:PGM} can be obtained from a further contraction of either~\eqref{eq:MGM} or~\eqref{eq:EGM} by scaling $Z_{ij}$ appropriately such that the overall scaling of $Z_{ij}$ and $Z_i$ match.

\subsection{Free Lie algebra embedding}

Similar to the embedding of the Maxwell algebra in a free Lie algebra~\cite{Gomis:2017cmt}, one has to find a consistent grading of the generators. In the Maxwell case one could take the grading where the Lorentz generators $M_{ab}$ had a level $\ell=0$, the translation generators $P_{a}$ were at level $\ell=1$ and the tensor generator $Z_{ab}$ then consistently at level $\ell=2$. As we shall see, there is no unique choice that works for all non-relativistic limits once extensions are taken into account.

While the ordinary Galilei algebra can be constructed as a contraction 
 of the Poincar\'e algebra using the first two lines of~\eqref{eq:contr}, 
it is not directly possible to construct a free Lie algebra extension of the Galilei algebra from a 
contraction of the free Lie algebra extension $\Minfty$ of the Poincar\'e algebra studied in~\cite{Gomis:2017cmt}. 
This can already be anticipated from investigations in~\cite{Bergshoeff:2015uaa,Bergshoeff:2016lwr,Barducci:2017mse,Hartong:2017bwq,Aviles:2018jzw} where it was necessary to extend the Poincar\'e or Maxwell algebra by additional abelian factors to obtain the correct contraction. We shall see below that one can obtain the different non-relativistic algebras~\eqref{eq:MGM}--\eqref{eq:PGM} by starting from different free Lie algebras with different level assignments.

In view of extensions of the Galilei algebra studied in~\cite{Bonanos:2008kr}, different level assignments appears more appropriate for embedding the various Galilean algebras in free Lie algebras. For instance, we might want to allow a non-trivial (central)\footnote{This extension is not strictly central in that $S_{ij}$ is a tensor under rotations but it commutes with all generators besides the rotation generators.}  extension 
\begin{align}
\lb G_i, G_j \rb = S_{ij} =- S_{ji}\,.
\end{align}
Leaving the boost generators at $\ell=0$ would also extend the $\ell=0$ algebra and thus not all extensions are captured by the free Lie algebra. In other words, we only would like to keep the rotation generator $M_{ij}$ at $\ell=0$ and move all other generators to positive levels. 
The simplest choice is to put the Hamiltonian $H$ and the boost generator $G_i$ at $\ell=1$ with all other level assignments induced by the structure of the algebra. This works well for the magnetic case~\eqref{eq:MGM} while for the electric case~\eqref{eq:EGM} one has to be careful in the case of the magnetic field $Z_{ij}$ that is not produced by any commutator---its level is fixed by the commutation relations~\eqref{eq:EGM}.
The fact that the spatial rotations are
considered at level zero is in agreement with the Eilenberg--Chevalley cohomology calculation done in~\cite{Bonanos:2008kr}.

An additional feature that we can use for the non-relativistic free Lie algebras is that there are two `types' of generators $H$ and $G_i$ at $\ell=1$ and we can introduce another (abelian) label that discriminates them and produces a second grading on the free Lie algebra. We shall call this label $m$ and $G_i$ is given $m=0$ while $H$ is given $m=1$, so that the $m$ counts basically the number of zero (time) indices compared to the relativistic case. 

\subsubsection{Magnetic Galilei Maxwell algebra}

As indicated above we consider the free Lie algebra based on the generators $H$ and $G_i$ at $\ell=1$ that transform under the spatial rotation group. We also assign $m=0$ to $G_i$ and $m=1$ to $H$. The resulting free Lie algebra is shown in Table~\ref{Tab:Free} for any space-time dimension $D$, where we also included the rotation generators $M_{ij}$ at $\ell=0$ and the Young tableaux are those of $\mf{gl}(D-1)$, {\it i.e.}, they contain traces when viewed as tensors of the rotation algebra $\mf{so}(D-1)$.

\setlength\arraycolsep{12pt}
\renewcommand\arraystretch{1.8}
\begin{table}
\begin{align*}
\begin{array}{c|c|c|c|c|c|c}
&\ell=0 &
\ell=1 &
\ell=2 &
\ell=3 &
\ell=4 &\cdots \\
\hline
m=0&  M_{ij}  
&   G_i 
&   S_{ij} 
&   Y_{ij,k}   
& \raisebox{-0.2\height}{\scalebox{0.5}{\yng(3,1)}}+\raisebox{-0.2\height}{\scalebox{0.5}{\yng(2,1,1)}}&\cdots
\\
m=1 &  
& H 
& P_i 
& B_{ij}\,, Z_{i,j}   
& \raisebox{-0.2\height}{\scalebox{0.5}{\yng(3)}} +2\,\raisebox{-0.2\height}{\scalebox{0.5}{\yng(2,1)}}+\raisebox{-0.2\height}{\scalebox{0.5}{\yng(1,1,1)}}&\cdots\\
m=2 &  
& 
&& Z_i 
&  \raisebox{-0.2\height}{\scalebox{0.5}{\yng(2)}}+2\, \raisebox{-0.2\height}{\scalebox{0.5}{\yng(1,1)}} &\cdots\\
m=3 &  
& 
& 
&  & \raisebox{-0.2\height}{\scalebox{0.5}{\yng(1)}} 
&\cdots
\end{array}
\end{align*}
\caption{\it The free Lie algebra extension of the magnetic Galilei Maxwell algebra, generated by $G_i$ and $H$
(together with the rotation generators $M_{ij}$) One of the two Young tableaux with two boxes in one column
at $(\ell,m)=(4,2)$ corresponds to the magnetic field generator $Z_{ij}$.}
\label{Tab:Free}
\end{table}

The table can be obtained by the algorithm described in~\cite{Cederwall:2015oua,Gomis:2017cmt,Gomis:2018xmo} and the level $m$ can be incorporated as an additional abelian charge. As the free Lie algebra is completely generated by everything at $\ell=1$, this means that we could refer to $\ell$ as the principal grading of the free Lie algebra such that $\ell$ counts how many commutators of the generating elements have been taken.

The free Lie algebra structure entails for example the following commutators of the basic generators:
\begin{subequations}
\label{GAF}
\begin{align}
\ell=1:&& H&\,,& G_i \,.\\
\ell=2:&& P_i &= \lb H, G_i \rb\,,&  S_{ij} &=\lb G_i, G_j \rb \,.\\
\ell=3:&& Z_i &= \lb H, P_i \rb\,, &B_{ij} &=\lb H, S_{ij} \rb\,,\nn\\
&& Z_{i,j} +\frac12 B_{ij} &=\lb G_i, P_j \rb \,,& Y_{ij,k} &= \lb  S_{ij} ,G_k\rb \,.\\
\label{eq:GAF4}
\ell=4:&& Z_{ij} &= \lb P_i, P_j \rb\,,& \ldots
\end{align}
\end{subequations}
The notation we use here and elsewhere is that groups of indices enclosed between commas are antisymmetric and belong to one column of a Young tableaux while a comma indicates the beginning of a new column with the associated Young irreducibility constraint. The translation to Young tableaux then is for instance
\begin{align}
Z_{ij} \leftrightarrow \yng(1,1) \,, \hspace{6mm}Z_{i,j} \leftrightarrow \yng(2) \,, \hspace{6mm}Y_{ij,k} \leftrightarrow \yng(2,1)\,.
\end{align}
The corresponding tensor symmetries are therefore
\begin{align}
Z_{ij} = -Z_{ji}\,,\quad Z_{i,j} = Z_{j,i} \,,\quad Y_{ij,k} = -Y_{ji,k}\quad\textrm{and the hook constraint}\quad Y_{[ij,k]} = 0\,.
\end{align}
At $\ell=4$ in~\eqref{eq:GAF4} we have only shown the commutator of the translation generators $\tilde{P}_i$ giving the magnetic field generator $Z_{ij} = -Z_{ji}$.

We note that the Jacobi identity fixes the anti-symmetric part in $\lb G_i, P_j\rb$ via
\begin{align}\label{eq:factor12}
 \lb G_{[i} , P_{j]} \rb = \lb G_{[i} ,\lb H, G_{j]} \rb\rb = \lb \lb G_{[i}, H \rb, G_{j]} \rb + \lb H, \lb G_i, G_j \rb\rb = -\lb G_{[i} ,\lb H, G_{j]} \rb\rb  + B_{ij}\,.
\end{align}

We can connect this free Lie algebra to the magnetic Galilei Maxwell algebra~\eqref{eq:MGM} by setting $S_{ij}=B_{ij}=Z_{i,j}=Y_{ij,k}=0$ and keeping only the electric field generator $Z_i$ at $\ell=3$ and the magnetic field generator $Z_{ij}$ at $\ell=4$.  
These restrictions at higher levels generate an infinite-dimensional ideal and therefore we can take the quotient of the Galilei free Lie algebra by this ideal. The commutation relations agree in the corresponding quotient.
We will further discuss such truncations in section \ref{sec:trunc}.

\subsubsection{Electric Galilei Maxwell algebra}

The electric Galilei Maxwell algebra does not fit directly into the same framework. We rather have to consider a different grading and even an enlarged set of generators of the free Lie algebra. This can be seen by looking at equation~\eqref{eq:EGM} and noticing that the magnetic field generators $Z_{ij}$ does not arise on the right-hand side of any commutator. Keeping $G_i$ and $H$ at level $\ell=1$ leads to $\ell=3$ for the electric field generator $Z_i$ as before. Therefore the consistency of the last commutator in the third column of~\eqref{eq:EGM} requires putting the magnetic field generators $Z_{ij}$ at level $\ell=2$---and it should appear as a new and independent generator of the free Lie algebra. Such a construction is possible if $\ell$ is not the principal grading as in the preceding section. Let us discuss this in a bit more generality.

Let us denote by  $X=(X_1,\ldots, X_n)$ the generating elements of the free Lie algebra $\mf{f}(X)$. The principal grading assigns level $\ell=1$ to each of them and keeps track of how many times the generating elements $X$ appear in a multiple commutator, {\it i.e.}\ level $\ell$ denotes an $\ell$-fold commutator of the $X_i$. 
This is only one possibility of grading a free Lie algebra with a given set of generators. The most general grading is obtained by keeping also track of which of the $X_i$ appears in a multiple commutator and this is captured for instance by formula (A.8) in~\cite{Gomis:2017cmt}.

The electric Galilei Maxwell algebra~\eqref{eq:EGM} requires the use of a different grading by assigning $\ell=2$ to the magnetic field generators $Z_{ij}$. In other words, we are considering the free Lie algebra on the generators
\begin{align}
X = \{ \underbrace{G_i\,,\quad H}_{\ell=1}\,,\quad\underbrace{Z_{ij}}_{\ell=2} \}\,,
\end{align}
where we have indicated the new $\ell$ level assignments. We denote the corresponding spaces of the generating set of the free Lie algebra by
\begin{align}
\mf{x}_1 = \left\langle G_i, \tilde{H} \right\rangle\,\quad 
\mf{x}_2 = \left\langle Z_{ij} \right\rangle\,.
\end{align}

The free Lie algebra $\mf{f}(X)$ then still is graded as $\mf{f}(X) = \bigoplus_{\ell>0} \mf{f}_\ell$ but the recursive algorithm obtained from specialising (A.8) in~\cite{Gomis:2017cmt} is different. It becomes
\begin{align}
\mf{f}_1 &= \mf{x}_1\,,\nn\\
\mf{f}_2 &= \ALT^2 \mf{f}_1 \oplus \mf{x}_2\,,\nn\\
\mf{f}_3 &= \mf{f}_2\otimes \mf{f}_1 \ominus \ALT^3 \mf{f}_1\,,\nn\\
\mf{f}_4 &= (\mf{f}_3\otimes \mf{f}_1\oplus \ALT^2\mf{f}_2)\ominus \mf{f}_2 \otimes \ALT^2\mf{f}_1 \oplus \ALT^4\mf{f}_1\,,\nn\\
\mf{f}_5 &= (\mf{f}_1\otimes \mf{f}_4 \oplus \mf{f}_2\otimes \mf{f}_3) \ominus (\mf{f}_3\otimes \ALT^2\mf{f}_1 \oplus  \mf{f}_1\otimes \ALT^2\mf{f}_2 ) \oplus \mf{f}_2\otimes \ALT^3\mf{f}_1 \ominus \ALT^5\mf{f}_1\,,
\end{align}
where the difference compared to the usual recursive algorithm is that now we have an extra piece from the additional generators $Z_{ij}$ contained in $\mf{x}_2$ contributing to $\mf{f}_2$.

Applying this formula, we obtain table~\ref{tab:MEB3} where we have also introduced the additional $m$-degree that distinguishes $H$ and $G_i$ and we have assigned $m$-degree $m=2$ to $Z_{ij}$. Note that at level $(\ell,m)=(3,2)$ we have two vectors: one associated to the Young tableau with one box and one
associated to the trace part of the hook
when decomposed into representations of $\mathfrak{so}(D-1)$.
If we want to recover the electric generator, we should impose
\begin{align}
[G_k, Z_{ij}]-2 \delta_{k[i} [H,P_{j]}]=0
\end{align}
in such a way 
we only
have one vector that we can call $Z_i$. In this way we get the
electric Galilei Maxwell algebra, if we also factor out the ideal generated by $S_{ij}$, 
everything else at $\ell=3$ and $[Z_{ij},Z_{kl}]$ at $(\ell,m)=(4,4)$.

The extended Maxwell exotic Bargmann algebra was already investigated in~\cite{Aviles:2018jzw} in $D=2+1$ dimensions and our analysis is in agreement with the results there when specialising to $D=2+1$, see Eq.~(3.25) in~\cite{Aviles:2018jzw}.
In  this case one can also simplify the calculation by using $Z_{ij} = \epsilon_{ij} Z$ where $Z$ is only a (pseudo-)scalar.

\setlength\arraycolsep{9pt}
\renewcommand\arraystretch{1.8}
\begin{table}
\begin{align*}
\begin{array}{c|c|c|c|c|c|c}
&\ell=0 &
\ell=1 &
\ell=2 &
\ell=3 &
\ell=4 &\cdots \\
\hline
m=0&  M_{ij}  
&   G_i 
&   S_{ij} 
&   Y_{ij,k}   
& \raisebox{-0.2\height}{\scalebox{0.5}{\yng(3,1)}}+\raisebox{-0.2\height}{\scalebox{0.5}{\yng(2,1,1)}}&\cdots
\\
m=1 &  
& H 
& P_i 
& B_{ij}\,, Z_{i,j}   
& \raisebox{-0.2\height}{\scalebox{0.5}{\yng(3)}} +2\,\raisebox{-0.2\height}{\scalebox{0.5}{\yng(2,1)}}+\raisebox{-0.2\height}{\scalebox{0.5}{\yng(1,1,1)}}&\cdots\\
m=2 &  
& 
& Z_{ij} 
& \raisebox{-0.2\height}{\scalebox{0.5}{\yng(1)}}+ \raisebox{-0.2\height}{\scalebox{0.5}{\yng(2,1)}}
+\raisebox{-0.2\height}{\scalebox{0.5}{\yng(1,1,1)}}  
&  \,\raisebox{-0.2\height}{\scalebox{0.5}{\yng(2)}}+2\,\raisebox{-0.2\height}{\scalebox{0.5}{\yng(1,1)}}
+\raisebox{-0.2\height}{\scalebox{0.5}{\yng(2,2)}}+\raisebox{-0.2\height}{\scalebox{0.5}{\yng(3,1)}}+2\,\raisebox{-0.2\height}{\scalebox{0.5}{\yng(2,1,1)}}
+\raisebox{-0.2\height}{\scalebox{0.5}{\yng(1,1,1,1)}}&\cdots\\
m=3 &  
& 
& 
&  \raisebox{-0.2\height}{\scalebox{0.5}{\yng(1,1)}}   
& \raisebox{-0.2\height}{\scalebox{0.5}{\yng(1)}}+2\,\raisebox{-0.2\height}{\scalebox{0.5}{\yng(2,1)}}
+2\,\raisebox{-0.2\height}{\scalebox{0.5}{\yng(1,1,1)}} &\cdots\\
m=4 &
&
&
&
& \raisebox{-0.2\height}{\scalebox{0.5}{\yng(1,1)}}+\raisebox{-0.2\height}{\scalebox{0.5}{\yng(2,1,1)}} &\cdots
\end{array}
\end{align*}
\caption{\it The free Lie algebra extension of the electric Maxwell Galilei algebra, generated by $G_i$, $H$ and $Z_{ij}$
(together with the rotation generators $M_{ij}$). The electric field generator $Z_{i}$ corresponds to a linear combination of the single box and the traceless part of the hook Young tableau at $(\ell,m)=(3,2)$.
}
\label{tab:MEB3}
\end{table}

\subsubsection{Pulse Galilei Maxwell algebra}

The pulse Galilei Maxwell algebra in~\eqref{eq:PGM} can also be obtained from the same free Lie algebra as the electric Galilei Maxwell algebra, generated by $G_i$, $H$ and $Z_{ij}$, by factoring out
the ideal generated by $S_{ij}$, everything at $\ell=3$ and $[Z_{ij},Z_{kl}]$ at $(\ell,m)=(4,4)$. One can see that it is better to include $Z_{ij}$ as an independent generator as it does not appear on the right-hand side of any commutator in~\eqref{eq:PGM}. In addition it commutes with all generators of~\eqref{eq:PGM} except for the rotation generators.

\subsection{Truncations} \label{sec:trunc}

By factoring out various ideals of the Galilei free Lie algebras one can obtain as quotients
the ordinary Galilei algebra, the electric and magnetic Galilei Maxwell 
algebras, but also other extensions of the Galilei algebra that have appeared in the literature.
If we set $S_{ij}=0$, then we have to set $Y_{ij,k}=B_{ij}=0$ as well, since these generators can be reached by acting on $S_{ij}$
with $G_{i}$ and $H$, respectively. Furthermore, not only $Y_{ij,k}$ has to be set to
zero, but also any other generator at $\ell\geq 2$ and $m=0$. In the magnetic case (the free Lie algebra generated by only $G_i$ and $H$) we are then left with
$Z_{i,j}$ and $Z_i$ at $\ell=3$. Keeping only $Z_i$ and setting $Z_{i,j}$ to zero gives the magnetic Galilei Maxwell algebra as we have discussed.
Keeping instead only the trace of $Z_{i,j}$ and setting $Z_i$ to zero gives the central Bargmann extension of the Galilei algebra.

We now consider the possibility of keeping $S_{ij}$ in the magnetic case. If we then still set to zero everything but the traces $B_{i}\sim \delta^{jk} Y_{ij,k}$ and $N \sim \delta^{ij} Z_{i,j}$ at $\ell=3$
then we get an extension of the Bargmann algebra, where the subalgebra at $m=0$ alternates between
two- and one-forms:
$M_{ij}$, $G_i$, $S_{ij}$, $B_i$ for $\ell=0,1,2,3$. Similarly, the $m=1$ subspace alternates between
singlets and one-forms:
$H$, $P_i$, $N$ for $\ell=1,2,3$. We can continue the alternating sequence at $m=1$
to $\ell=4$ by including $T_i=[N, G_i]$ and setting everything else at $\ell=4$ to zero, as well as everything at higher levels $\ell$.
We then obtain a Lie algebra with the following nonzero commutation relations (not involving rotations):
\begin{subequations}
\label{Obers}
\begin{align}
[H,G_i]&=P_i\,,\\
[P_i,G_j]&=\delta_{ij}N\,,\\
[N,G_i]&=T_i\,,\\
[H,B_i]&=T_i\,,\\
[S_{ij},P_k]&=2\delta_{k[i}T_{j]}\,,\\
[G_i,G_j]&=-S_{ij}\,,\\
[S_{ij},G_k]&=2\delta_{k[i}B_{j]}\,.
\end{align}
\end{subequations}
This Lie algebra
has recently been considered as underlying Newton gravity in the same way as the Poincar\'e algebra underlies general relativity \cite{Hansen:2018ofj}.
By subsequently factoring out the appropriate ideals, this alternating pattern can be continued to an arbitrary number of diagonal levels,
and also to infinity. The algebra in \cite{Hansen:2018ofj} was extended to $(\ell,m)=(4,0)$ and $(\ell,m)=(5,1)$ in \cite{Ozdemir:2019orp} (for $D=3$, in order to construct a corresponding Chern--Simons action) and to infinity in 
\cite{Hansen:2019vqf}. 
These algebras are shown in table \ref{Tab:Ozdemir}, where we have introduced a new ``diagonal'' grading, given by $n=\ell-m$,
and in table \ref{Tab:Obers2}.

\begin{table}[h]
\begin{align*}
\xymatrix@!0@C=2.0cm{
\ar@{-}[]+<1.1cm,1em>;[ddd]+<1.1cm,-1em> \ar@{-}[]+<-0.8cm,-1em>;[rrrrrrr]+<0.6cm,-1em>&
\ell=0 \ar@{-}[]+<1.1cm,1em>;[ddd]+<1.1cm,-1em>&
\ell=1\ar@{-}[]+<1.1cm,1em>;[ddd]+<1.1cm,-1em>&
\ell=2\ar@{-}[]+<1.1cm,1em>;[ddd]+<1.1cm,-1em>&
\ell=3\ar@{-}[]+<1.1cm,1em>;[ddd]+<1.1cm,-1em>&\ell=4\ar@{-}[]+<1.1cm,1em>;[ddd]+<1.1cm,-1em>&\ell=5\ar@{-}[]+<1.1cm,1em>;[ddd]+<1.1cm,-1em> &\cdots\\ 
m=0&M_{ij} 
&     G_i 
&   S_{ij}  
&   B_i  
\ar@{-}@[blue][dr] & Z_{ij} & \cdots & \cdots
\\
m=1 &  
& H \ar@{-}@[blue][ul] 
& P_i \ar@{-}@[blue]@[blue][ul]  
&  N \ar@{-}@[blue][ul]& T_i \ar@{-}@[blue][ul] & \ar@{-}@[blue][ul] Y & \cdots\\
& 
& & *+[F-:blue][blue]{ n=0}\ar@{-}@[blue][ul] & *+[F-:blue][blue]{ n=1}*\frm{-}\ar@{-}@[blue][ul] & *+[F-:blue][blue]{ n=2}\ar@{-}@[blue][ul]
& *+[F-:blue][blue]{ n=3}\ar@{-}@[blue][ul] &*+[F-:blue][blue]{ n=4}\ar@{-}@[blue][ul]
}
\end{align*}
\caption{\it The extension of the algebra in \cite{Hansen:2018ofj} with one more diagonal level \cite{Ozdemir:2019orp}.}
\label{Tab:Ozdemir}
\end{table}

\begin{table}[h]
\begin{align*}
\xymatrix@!0@C=2.0cm{
\ar@{-}[]+<1.1cm,1em>;[ddd]+<1.1cm,-1em> \ar@{-}[]+<-0.8cm,-1em>;[rrrrrrr]+<0.6cm,-1em>&
\ell=0 \ar@{-}[]+<1.1cm,1em>;[ddd]+<1.1cm,-1em>&
\ell=1\ar@{-}[]+<1.1cm,1em>;[ddd]+<1.1cm,-1em>&
\ell=2\ar@{-}[]+<1.1cm,1em>;[ddd]+<1.1cm,-1em>&
\ell=3\ar@{-}[]+<1.1cm,1em>;[ddd]+<1.1cm,-1em>&\ell=4\ar@{-}[]+<1.1cm,1em>;[ddd]+<1.1cm,-1em>& \ell=5\ar@{-}[]+<1.1cm,1em>;[ddd]+<1.1cm,-1em> & \cdots\\ 
m=0&J_{ij}{}^{(0)} 
&     B_i{}^{(0)} 
&   J_{ij}{}^{(1)}  
&   B_i{}^{(1)}  
& J_{ij}{}^{(2)} & \cdots & \cdots
\\ \ar@{-}[]+<-0.8cm,-1em>;[rrrrrrr]+<0.6cm,-1em>
m=1 &  
& H^{(0)} 
& P_i^{(0)} 
&  H^{(1)} 
& P_i^{(1)} & H^{(2)} & \cdots\\ 
& \mathcal{J}_{ab}{}^0
& \mathcal{P}_{a}{}^1& 
\mathcal{J}_{ab}{}^2 
& \mathcal{P}_{a}{}^3 
& \mathcal{J}_{ab}{}^4
& \mathcal{P}_{a}{}^5 & \cdots 
}
\end{align*}
\caption{\it The generators named as in \cite{Hansen:2019vqf} in the first two rows.
The generators of the affine algebra are given in the last row, where the superscript is just the level $\ell$, and decomposes into
those in the rows above under the $\mathfrak{so}(D-1)$ subalgebra of $\mathfrak{so}(1,D-1)$.}
\label{Tab:Obers2}
\end{table}

We can thus obtain the algebra in \cite{Hansen:2019vqf} by factoring out infinitely many ideals of the free Lie algebra extension of the
Galilei Maxwell algebra, but these ideals have to
be determined recursively level by level such that we are left with only the desired generators. In the next section we will see that
the algebra can also be obtained more directly by factoring out only one ideal, corresponding to one of the Serre relations in an affine Kac--Moody algebra,
and then performing a contraction.

\section{Construction from affine Kac--Moody algebras}
\label{sec:AKM}

Consider the free Lie algebra where the generators $\mathcal{P}_a$ are vectors under $\mathfrak{so}(1,D-1)$.
As shown in \cite{Gomis:2017cmt} this gives an infinite-dimensional extension of the (relativistic) Maxwell algebra in $D$ dimensions. 
From the basic generators $\mathcal{P}_a$ at level $\ell=1$ we get
\begin{align}
\lb \mathcal{P}_a, \mathcal{P}_b \rb = \mathcal{Z}_{ab}\,,\quad \lb \mathcal{Z}_{ab}, \mathcal{P}_c\rb = \mathcal{Y}_{ab,c}
\end{align}
at level $\ell=2$ and $\ell=3$, respectively. 
The generator $\mathcal{Z}_{ab}$ is antisymmetric and $\mathcal{Y}_{ab,c}$ has the hook symmetry under $\mathfrak{gl}(D)$.
By splitting the indices according to $a=(0,i)$ into time and space we get the free 
Lie algebra extension of the magnetic Galilei Maxwell algebra
above generated by the $\mathfrak{so}(D-1)$ vectors $G_i$ and scalars $H$.
From this point of view the level $m$ counts how many times the index $0$ appears in the tensors.
At the first three levels we have explicitly
\begin{subequations}
\begin{align}
\ell=1:&&&H \lra \mathcal{P}_0\,,\quad G_i \lra \mathcal{P}_i\\
\ell=2:&&& P_i \lra \mathcal{Z}_{0i}=-\mathcal{Z}_{i0}\,,\quad S_{ij} \lra \mathcal{Z}_{ij}\\
\ell=3:&&& Z_i \lra \mathcal{Y}_{i0,0}\,,\quad B_{ij} \lra -\mathcal{Y}_{ij,0}\,,\nn\\
&&& Z_{i,j} \lra -\mathcal{Y}_{0(i,j)} \,,\quad Y_{ij,k} \lra \mathcal{Y}_{ij,k}\,.
\end{align}
\end{subequations}
Note that $Y_{0[i,j]}$ corresponds to $\frac12 B_{ij}$ because of the Young irreducibility of $\mathcal{Y}_{ab,c}$, see~\eqref{eq:factor12}.

As we will see, by factoring out the ideal generated by
$\eta^{bc}\mathcal{Y}_{ab,c}=\eta^{00}\mathcal{Y}_{a0,0}+\eta^{ij}\mathcal{Y}_{ai,j}$ we get, together with the rotation generators $M_{ij}$
at $\ell=0$, a Lie algebra which has the same structure as the one in \cite{Hansen:2019vqf}, shown in table \ref{Tab:Obers2}. However, as we will see,
the commutation
relations will be different, until we also perform a contraction.
This is a subalgebra of a (twisted or untwisted) affine Lie algebra which is an extension of 
$\mathfrak{so}(1,D-1)$. In order to describe it we first need to consider the cases of odd and even $D$ separately.

\begin{figure}[t!]
\centering
\begin{picture}(227,87.2)
\thicklines
\put(10,20){\circle{7}}
\put(13.5,21){\line(1,0){30.2}}
\put(13.5,19){\line(1,0){30.2}}
\put(13.5,20){\line(1,1){5}}
\put(13.5,20){\line(1,-1){5}}
\put(47,20){\circle{7}}
\put(184,20){\circle{7}}
\put(184,23.5){\line(0,1){30.2}}
\put(184,57.2){\circle{7}}
\put(187.7,20){\line(1,0){30.2}}
\put(221.7,20){\circle{7}}
\put(50,20){\line(1,0){30.2}}
\put(180.4,20){\line(-1,0){30.2}}
\put(84.1,20){\circle{7}}
\put(146.3,20){\circle{7}}
\multiput(87.5,20)(22,0){3}{\line(1,0){11}}
\put(7.5,5){$0$}
\put(44,5){$1$}
\put(210,5){$r-1$}
\put(195,55){$r$}
\end{picture}
\begin{picture}(227,50)
\thicklines
\put(10,20){\circle{7}}
\put(14.5,21){\line(1,0){29.2}}
\put(14.5,19){\line(1,0){29.2}}
\put(47,20){\circle{7}}
\put(13.5,20){\line(1,1){5}}
\put(13.5,20){\line(1,-1){5}}
\put(218.2,20){\line(-1,1){5}}
\put(218.2,20){\line(-1,-1){5}}
\put(184,20){\circle{7}}
\put(187.7,21){\line(1,0){29.2}}
\put(187.7,19){\line(1,0){29.2}}
\put(221.7,20){\circle{7}}
\put(50,20){\line(1,0){30.2}}
\put(180.4,20){\line(-1,0){30.2}}
\put(84.1,20){\circle{7}}
\put(146.3,20){\circle{7}}
\multiput(87.5,20)(22,0){3}{\line(1,0){11}}
\put(7.5,5){$0$}
\put(44,5){$1$}
\put(220,5){$r$}
\end{picture}
\caption{
\it Dynkin diagrams of $B_r^{(1)}$ (upper) and $D_{r+1}^{(2)}$ (lower).}\label{fig:Br2}
\end{figure}
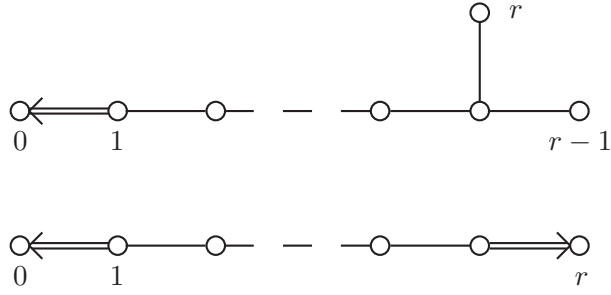

The Lorentz algebra $\mathfrak{so}(1,D-1)$ is a real form of $B_r$ if $D$ is odd, $D-1=2r$, and of $D_r$ if $D$ is even, $D-1=2r-1$. 
Both $B_r$ and $D_r$ can be extended to affine Lie algebras by adding 
a node $0$ to the Dynkin diagram, attached to node $1$ with a double line pointing outwards, as shown in figure \ref{fig:Br2}.
The resulting affine Lie algebra is here denoted
$B_r^{(1)}$ if $D=2r$ and 
$D_{r+1}^{(2)}$ if $D-1=2r$, following the standard terminology~\cite{Kac}. The algebra $D_{r+1}^{(2)}$ is a {\it twisted} affine algebra, whereas $B_r^{(1)}$ is untwisted \cite{Kac}.\footnote{The untwisted affine algebra $B_r^{(1)}$ is built from the horizontal subalgebra $B_r$ that is obtained by removing node $r$ from its Dynkin diagram and considering the associated loop algebra, together with a central extension and derivation element. The twisted affine algebra $D_{r+1}^{(2)}$ is constructed from the loop algebra of maps from the circle into the algebra $D_{r+1}$ that differ by an application of the order-2 diagram automorphism 
exchanging the two `spinor nodes'
when going around the circle. The subalgebra of the `horizontal' $D_{r+1}$ that is invariant under this automorphism is $B_r$ and its diagram is obtained by removing node $r$ (or $0$) from the diagram of $D_{r+1}^{(2)}$. 

}

The Cartan matrix is
\setlength\arraycolsep{10pt}
\renewcommand\arraystretch{1}
\begin{align}
A_{ij}=
\begin{pmatrix}
2 & -2 & 0 &  \cdots & 0 & 0& 0\\ 
-1 & 2 & -1 &  \cdots & 0 & 0& 0\\ 
0 & -1 & 2&  \cdots & 0 & 0& 0\\ 
\vdots &  \vdots &  \vdots &  \ddots & \vdots & \vdots& \vdots  \\
0 & 0 & 0 & \cdots & 2 & -1 & -1\\
0 & 0 & 0 & \cdots & -1 & 2 & 0\\
0 & 0 & 0 & \cdots & -1 & 0 & 2
\end{pmatrix}
\end{align} 
for $B_r{}^{(1)}$
and
\setlength\arraycolsep{10pt}
\renewcommand\arraystretch{1}
\begin{align}
A_{ij}=
\begin{pmatrix}
2 & -2 & 0 &  \cdots & 0 & 0 & 0\\ 
-1 & 2 & -1 &  \cdots & 0 & 0 & 0\\ 
0 & -1 & 2&  \cdots & 0 & 0 & 0\\ 
\vdots &  \vdots &  \vdots &  \ddots & \vdots & \vdots & \vdots  \\
0 & 0 & 0 & \cdots &2& -1 & 0\\
0 & 0 & 0 & \cdots &-1& 2 & -1\\
0 & 0 & 0 & \cdots &0& -2 & 2
\end{pmatrix}
\end{align} 
for $D_{r+1}{}^{(2)}$. Here, $i,j=0,1,\ldots,r$. The affine algebra (either $B_r{}^{(1)}$ or $D_{r+1}{}^{(2)}$)
is generated by $3(r+1)$ elements $e_i,f_i,h_i$ modulo the
Chevalley--Serre relations
\begin{align}
[h_i,e_j]&=A_{ij}e_j\,,& [h_i,f_j]&=-A_{ij}f_j\,,& [e_i,f_j]&=\delta_{ij}h_j\,, & [h_i,h_j]&=0\,,
\end{align}
\begin{align} \label{serre}
({\rm ad}\, e_i)^{1-A_{ij}}(e_j)=({\rm ad}\, f_i)^{1-A_{ij}}(f_j)=0\,.
\end{align}
It can be given a $\mathbb{Z}$-grading where the level $\ell$ is the number of $e_0$ generators if $\ell\geq 0$ (and otherwise $-\ell$ is the number of
$f_0$ generators). We are here interested in the (Borel) subalgebra generated by all generators but $f_0$, which thus only has non-negative levels $\ell$.

The Serre relations (\ref{serre}) associated to the pair of nodes $0$ and $1$ 
(in both affine algebras) are
\begin{align}
[e_1,[e_1,e_0]]&=0\,,&  [e_0,[e_0,[e_0,e_1]]]&=0\,.
\end{align}
However, $[e_1,e_0]$ and $[e_0,[e_0,e_1]]$ are nonzero elements at level 1 and 2, respectively.
When we decompose the algebra with respect to node $0$ we have $e_0$ as a lowest weight vector at level $1$ since $[f_i,e_0]=0$ for 
$i=1,2,\ldots,r$ and the associated $r$ Dynkin labels (of the corresponding {\it highest} weight representation) are $[1,0,0,\ldots,0]$ since
\begin{align}
[h_1,e_0]&=-1\,,& [h_2,e_0]&=[h_3,e_0]=\cdots = [h_r,e_0]=0\,.
\end{align}
We then get the algebra from the free Lie algebra generated by this representation by factoring out the ideal generated by the Serre relation $[e_0,[e_0,[e_0,e_1]]]=0$. Since it appears at level $3$
we have the full content of the free Lie algebra at level $2$, that is, $[0,1,0,\ldots,0]$ (the antisymmetric product of two $[1,0,0,\ldots,0]$'s). The corresponding lowest weight vector is $[e_0,[e_0,e_1]]$. The Dynkin labels associated to $[e_0,[e_0,[e_0,e_1]]]=0$ is $[1,1,0,\ldots,0]$, so this representation has to be removed from the content of the free Lie algebra at level 3, and also all representations at higher levels in the ideal that it generates. Thus only $[1,0,0,\ldots,0]$
remains at level 3, and it turns out that this pattern of alternating $[1,0,0,\ldots,0]$ and $[0,1,0,\ldots,0]$ continues to all levels.
This can be checked with the SimpLie software \cite{SimpLie}.
We can thus introduce generators $\mathcal{J}_{ab}{}^\ell$ at all even levels $\ell$ and $\mathcal{P}_{a}{}^\ell$ at all odd levels $\ell$.
We get the commutation relations  
\begin{subequations}
\label{JP}
\begin{align}
[\mathcal{J}_{ab}{}^{2k},\mathcal{J}_{cd}{}^{2k'}]&=4\eta_{[c[b}\mathcal{J}_{a]d]}{}^{2(k+k')}\,,\\
[\mathcal{J}_{ab}{}^{2k},\mathcal{P}_c{}^{2k'+1}]&=2\eta_{c[b}\mathcal{P}_{a]}{}^{2(k+k')+1}\,,\\
[\mathcal{P}_a{}^{2k+1},\mathcal{P}_b{}^{2k'+1}]&=\mathcal{J}_{ab}{}^{2(k+k'+1)}\,,
\end{align}
\end{subequations}
where the superscript is the level $\ell$.

We note that this subalgebra of the affine Kac--Moody algebra is an infinite-dimensional extension of the $\mathfrak{B}_k$ algebras obtained by Lie algebra expansions in \cite{Salgado:2014qqa} and thus it could be denoted $\mathfrak{B}_\infty$. For a given $k$, the finite-dimensional Lie algebra $\mathfrak{B}_k$ is obtained by factoring out the ideal generated by everything at level $\ell=k-1$. In $D=3$ space-time dimensions infinite extensions of the $\mathfrak{B}_k$ algebras have also been considered in
\cite{Caroca:2017onr}.

Next we decompose the $\mathfrak{so}(1,D-1)$ tensors into $\mathfrak{so}(D-1)$ tensors by setting
\begin{align}
\label{eq:br1}
J_{ij}{}^{(k)}&= \mathcal{J}_{ij}{}^{2k}\,, & B_i{}^{(k)} &= \mathcal{P}_i{}^{2k+1}\,,\nn\\
P_{i}{}^{(k-1)} &= \mathcal{J}_{0i}{}^{2k}\,, &  H^{(k)} &=\mathcal{P}_0{}^{2k+1}\,.
\end{align}
The algebra~\eqref{JP} in this basis of generalized boost, spatial momenta and Hamiltonian generators becomes
\begin{subequations}
\label{JPdecomp}
\begin{align}
[J_{ij}{}^{(m)},J_{kl}{}^{(n)}]&=4 \delta_{[k[j} J_{i]l]}{}^{(m+n)}\,,\\ 
[J_{ij}{}^{(m)},H^{(n)}]&=0\,,\\ 
[J_{ij}{}^{(m)},B_k{}^{(n)}]&=2\delta_{k[j}B_{i]}{}^{(m+n)}\,,\\ 
[J_{ij}{}^{(m)},P_k{}^{(n)}]&=2\delta_{k[j}P_{i]}{}^{(m+n)}\,,\\
[H^{(m)},B_i{}^{(n)}]&=P_i{}^{(m+n)}\,,\\ 
[P_i{}^{(m)},B_j{}^{(n)}]&=\delta_{ij}H^{(m+n+1)}\,,\\ 
[B_i{}^{(m)},B_j{}^{(n)}]&=J_{ij}{}^{(m+n+1)}\,\label{BBJ}\\ 
[H^{(m)},P_i{}^{(n)}]&=-B_i{}^{(m+n+1)}\,\label{HPB}\\ 
[P_i{}^{(m)},P_j{}^{(n)}]&=J_{ij}{}^{(m+n+2)},\label{PPJ}\\ 
[H{}^{(m)},H{}^{(n)}]&=0\,.
\end{align}
\end{subequations}
On top of the $\mathbb{Z}$-grading by $\ell$ inherited from~\eqref{JP}, we see that this algebra admits a $\mathbb{Z}_2$-grading, where we assign (as before)
$\mathbb{Z}_2$-degree $m=0$ to those generators of~\eqref{eq:br1} that do not carry a $0$ space-time index (in the first line) while we assign $m=1$ to those that do (in the second line), see table \ref{Tab:Obers2}.
The $\mathbb{Z}_2$-grading means that the commutator of two generators at $m=1$ gives a generator at $m=0$ (not something at $m=2$,
which would have to be zero, since there are only generators at $m=0$ and $m=1$).

This algebra generalizes the Poincar\'e algebra. In order to turn it into a non-relativistic algebra we perform the contraction
\begin{align}
\tilde J_{ij} &= J_{ij}\,, & \tilde B_i &=  B_i\,,\nn\\
\tilde P_{i} &= \frac1{\omega}P_i\,, & \tilde H &=   \frac1{\omega}H\,, \label{eq:contr2}
\end{align}
where we have suppressed the superscript since it does not play any role. In the limit $\omega \to \infty$,
the commutators (\ref{HPB}) and (\ref{PPJ}) will then vanish, and we end up with the commutation relations (1)
in \cite{Hansen:2019vqf} (up to a minus sign in the last equation there, which is  (\ref{BBJ}) here, because of different conventions for the Lorentz algebra in 
\cite{Hansen:2019vqf}). Note that this converts the $(\mathbb{Z}\times \mathbb{Z}_2)$-grading to a $(\mathbb{Z}\times \mathbb{Z})$-grading as now the commutators leading to $m=2$ vanish and thus the grading can be extended trivially to $\mathbb{Z}$.

The generators obtained in this process are exactly the ones shown in table~\ref{Tab:Ozdemir}. We note that the diagonal $n$-level shown rearranges the generators into repeating copies of $V_0=\langle M_{ij}, H\rangle$ and $V_1=\langle G_i, P_i\rangle$ and the expansion in $n$ is precisely the Lie algebra expansion used in~\cite{Bergshoeff:2019ctr}. Thus the Lie algebra expansion method can also be understood in terms of (affine) Kac--Moody algebras by adding nodes to the initial symmetry algebras as explained in this section. 
Furthermore, the algebras of~\cite{Hansen:2019vqf,Ozdemir:2019orp,Bergshoeff:2019ctr} can be understood as contractions of the algebras $\mathfrak{B}_k$ of~\cite{Salgado:2014qqa} in light of the comments below~\eqref{JP}.

\section{String Galilei algebra}
\label{sec:SGA}

There is a natural extension of the Galilei algebra that is called the string Galilei algebra
~\cite{Brugues:2004an, Brugues:2006yd} see also \cite{Barducci:2019fjc}.
In the same way that the Galilei algebra has a special direction $0$ corresponding to the longitudinal direction of the world-line of a particle, one can consider the $(p+1)$-dimensional world-volume of an extended $p$-dimensional object and treat 
all the longitudinal directions differently from the transverse ones.\footnote{In general,
there are $p+1$ possible contractions of the relativistic algebra \cite{Batlle:2016iel}. Here, we consider
the most symmetric case among the longitudinal variables by scaling them all in the same way.}

The simplest case would be $p=1$ corresponding to a string, but higher branes are also possible. 
We use the notation that $\alpha=0,\ldots, p$ denotes the longitudinal indices and continue to use $i$ for the transverse directions. The relativistic Lorentz and translation generators decompose under $SO(1,D-1)\to SO(1,p) \times SO( D-1-p)$ with this convention as
\begin{align}
M_{ab} &\to M_{ij},\, G_{i\alpha} ,\, M_{\alpha\beta}\,,\nn\\
P_a &\to P_i ,\, H_\alpha\,.
\end{align}
The new feature compared to the point particle is now the rotation generator 
 $M_{\alpha\beta}$ in the world-volume and that the Hamiltonian $H$ is replaced by a family of `Hamiltonians' $H_\alpha$ with a similar additional index for the boosts $G_{i\alpha}$. 
In the above equation we already have made the transition to the non-relativistic limit by performing a contraction analogous to~\eqref{eq:contr}:
 \begin{align}
\tilde{M}_{ij} &= M_{ij}\,, & \tilde{M}_{\alpha\beta} &= M_{\alpha\beta}\,,& \tilde{G}_{i\alpha} &= \frac1{\omega} M_{i\alpha}\,,\nn\\
\tilde{H}_\alpha &= \omega P_\alpha\,, & \tilde{P}_i &= P_i\,.
\end{align}

The non-trivial algebra of these generators obtained by the contraction of the Poincar\'e algebra is
\begin{align}
\lb  H_\alpha, G_{i\beta} \rb = -\eta_{\alpha\beta} P_i\,,\quad 
\lb M_{ij} , G_{k\alpha} \rb =-2 \delta_{k[i} G_{j]\alpha} \,, \quad
\lb M_{\alpha\beta} , G_{i\gamma} \rb = 2 \eta_{\gamma[\alpha} G_{i\beta]}\,.
\end{align}
Here, $\eta_{\alpha\beta}=(-+\cdots +)$ is the flat Minkowski metric along the world-volume and $\delta_{ij}$ is the flat Euclidean metric in the transverse space. 

We can construct a free Lie algebra extension of this that will generalise the magnetic Galilei Maxwell algebra~\eqref{eq:MGM} by taking as the generating set the Hamiltonians $H_\alpha$ and the boosts $G_{i\alpha}$ on $\ell=1$. This produces Table~\ref{tab:FreeBrane}.

\setlength\arraycolsep{8pt}
\renewcommand\arraystretch{1.8}
\begin{table}
\begin{align*}
\begin{array}{c|c|c|c|c|c}
&\ell=0 &
\ell=1 &
\ell=2 &
\ell=3 &
\cdots \\
\hline
m=0& \scalebox{.8}{$\lp\,\raisebox{-0.2\height}{\scalebox{0.5}{\yng(1,1)}}\,,\bullet\rp + \lp\bullet,\raisebox{-0.2\height}{\scalebox{0.5}{\yng(1,1)}}\, \rp$} 
&   \scalebox{.8}{$\lp\,  \raisebox{-0.1\height}{\scalebox{0.5}{\yng(1)}}\, ,\raisebox{-0.1\height}{\scalebox{0.5}{\yng(1)}} \,\rp$}
&  \scalebox{.8}{$ \lp\, \raisebox{-0.2\height}{\scalebox{0.5}{\yng(1,1)}}  \,,\,\raisebox{-0.1\height}{\scalebox{0.5}{\yng(2)}} \,\rp +  \lp\, \raisebox{-0.1\height}{\scalebox{0.5}{\yng(2)}}  \,,\,\raisebox{-0.2\height}{\scalebox{0.5}{\yng(1,1)}} \,\rp$}
& \scalebox{.8}{$\begin{array}{c}  \lp \, \raisebox{-0.1\height}{\scalebox{0.5}{\yng(3)}}  \,,\raisebox{-0.2\height}{\scalebox{0.5}{\yng(2,1)}} \rp +\lp \, \raisebox{-0.2\height}{\scalebox{0.5}{\yng(2,1)}}  \,,\raisebox{-0.1\height}{\scalebox{0.5}{\yng(3)}} \rp +\lp \, \raisebox{-0.3\height}{\scalebox{0.5}{\yng(1,1,1)}}  \,,\raisebox{-0.2\height}{\scalebox{0.5}{\yng(2,1)}} \rp \\+\lp \, \raisebox{-0.2\height}{\scalebox{0.5}{\yng(2,1)}}  \,,\raisebox{-0.3\height}{\scalebox{0.5}{\yng(1,1,1)}} \rp+\lp \, \raisebox{-0.2\height}{\scalebox{0.5}{\yng(2,1)}}  \,,\raisebox{-0.2\height}{\scalebox{0.5}{\yng(2,1)}} \rp\end{array}$}
&\cdots
\\
m=1 &  
& \scalebox{.8}{$\lp\bullet,  \raisebox{-0.1\height}{\scalebox{0.5}{\yng(1)}}\,\rp$}
& \scalebox{.8}{$\lp\,\raisebox{-0.1\height}{\scalebox{0.5}{\yng(1)}} \,,\,\raisebox{-0.1\height}{\scalebox{0.5}{\yng(2)}} \,\rp + \lp\,\raisebox{-0.1\height}{\scalebox{0.5}{\yng(1)}} \,,\,\raisebox{-0.2\height}{\scalebox{0.5}{\yng(1,1)}} \,\rp$}
& \scalebox{.8}{$\begin{array}{c}\lp \raisebox{-0.2\height}{\scalebox{0.5}{\yng(1,1)}},\raisebox{-0.1\height}{\scalebox{0.5}{\yng(3)}} \rp +  2\times\lp \raisebox{-0.2\height}{\scalebox{0.5}{\yng(1,1)}},\raisebox{-0.2\height}{\scalebox{0.5}{\yng(2,1)}} \rp + \lp \raisebox{-0.2\height}{\scalebox{0.5}{\yng(1,1)}},\raisebox{-0.3\height}{\scalebox{0.5}{\yng(1,1,1)}} \rp \\+ \lp \raisebox{-0.1\height}{\scalebox{0.5}{\yng(2)}},\raisebox{-0.3\height}{\scalebox{0.5}{\yng(1,1,1)}} \rp + 2\times\lp \raisebox{-0.1\height}{\scalebox{0.5}{\yng(2)}},\raisebox{-0.2\height}{\scalebox{0.5}{\yng(2,1)}} \rp+ \lp \raisebox{-0.1\height}{\scalebox{0.5}{\yng(2)}},\raisebox{-0.1\height}{\scalebox{0.5}{\yng(3)}} \rp\end{array}$}
&\cdots\\
m=2 &  
& 
&\scalebox{.8}{$\lp\bullet,  \raisebox{-0.2\height}{\scalebox{0.5}{\yng(1,1)}}\,\rp$}
&\scalebox{.8}{$\lp \raisebox{-0.1\height}{\scalebox{0.5}{\yng(1)}},\raisebox{-0.3\height}{\scalebox{0.5}{\yng(1,1,1)}} \rp +  \lp \raisebox{-0.1\height}{\scalebox{0.5}{\yng(1)}},\raisebox{-0.1\height}{\scalebox{0.5}{\yng(3)}} \rp +    2\times \lp \raisebox{-0.2\height}{\scalebox{0.5}{\yng(1)}},\raisebox{-0.2\height}{\scalebox{0.5}{\yng(2,1)}} \rp $}
&\cdots
\\
m=3 &  
& 
& 
&\scalebox{.8}{$\lp\bullet,\raisebox{-0.2\height}{\scalebox{0.5}{\yng(2,1)}} \rp   $}
&\cdots
\end{array}
\end{align*}
\caption{\it The free Lie algebra for the `magnetic' string Galilei algebra. As objects now transform under $SO(D-p-1) \times SO(1,p)$ we list the representations as pairs of Young tableaux (of the corresponding linear group). The first entry in each pair refers to the transverse rotation group.}
\label{tab:FreeBrane}
\end{table}

In that table, we have only shown things up to level $\ell=2$, but the structure can be easily generalised. The corresponding commutation in the free Lie algebra going to $\ell=2$ are
\begin{align}
m=0:&& \lb G_{i\alpha}, G_{j\beta} \rb &=  S_{ij \alpha,\beta} + S_{i,j \alpha\beta}\,,\\
m=1:&& \lb H_\alpha , G_{i \beta} \rb &= P_{i \alpha,\beta} + P_{i \alpha\beta}\,,\\
m=2:&& \lb H_{\alpha} , H_\beta \rb &= W_{\alpha\beta}\,.
\end{align}
We use the comma labelling convention for Young tableaux for both of the groups $SO(D-p-1)$ and $SO(1,p)$.
The usual string Galilei momentum generator $P_i$ is contained in these commutation relations as the trace $P_i = -\frac{1}{p+1} \eta^{\alpha\beta} P_{i \alpha,\beta}$. Moreover, the transverse trace of $S_{i,j\alpha\beta}$ gives $Z_{\alpha\beta} = -\frac{1}{D-p-1} \delta^{ij} S_{i,j\alpha\beta}$ which is an extension that appears as
\begin{align}
\lb G_{i\alpha}, G_{j\beta} \rb &=  \delta_{ij} Z_{\alpha\beta} + \ldots
\end{align}
in the commutator of two boost generators and has been studied before in the literature~\cite{Brugues:2004an, Brugues:2006yd}. The generator $Z_{\alpha\beta}$ appears at $(\ell,m)=(2,0)$ in the free Lie algebra.

For level $\ell=3$ we do not present complete commutation relations but only some that are relevant for comparing with the usual string Galilei algebra and its Maxwell extensions. Specifically, we note that
\begin{align}
\lb P_i, G_{j\alpha} \rb = \delta_{ij} Z_\alpha\,,\quad
\lb H_\alpha, Z_{\beta\gamma} \rb = \eta_{\alpha[\beta} Z_{\gamma]} + \ldots\,.
\end{align}
The new generator $Z_\alpha$ arises at  $(\ell,m)=(3,1)$ in the free Lie algebra and has appeared in the literature before in a different context~\cite{Brugues:2004an, Brugues:2006yd}. In fact, there are three different possible occurrences of such a tensor structure at $(\ell,m)=(3,1)$ by taking traces over the symmetric transverse indices and traces of the longitudinal representations (that is either a hook or a completely symmetric representation).

In summary, we see that the free Lie algebra approach has more than ample room to accommodate the different non-relativistic kinematic algebras that have appeared in a particle or brane context.

\end{document}